\documentclass[prx,twocolumn,superscriptaddress,showpacs]{revtex4}
\usepackage[dvips]{graphicx}
\usepackage{latexsym}
\usepackage{amsmath}
\usepackage{amsfonts}
\usepackage{amssymb}
\usepackage{bm}
\usepackage{color}
\usepackage{hhline}

\begin{document}
\newcommand{\fig}[2]{\includegraphics[width=#1]{#2}}
\newcommand{{\vhf}}{\chi^\text{v}_f}
\newcommand{{\vhd}}{\chi^\text{v}_d}
\newcommand{{\vpd}}{\Delta^\text{v}_d}
\newcommand{{\ved}}{\epsilon^\text{v}_d}
\newcommand{{\vved}}{\varepsilon^\text{v}_d}
\newcommand{{\bk}}{{\bf k}}
\newcommand{{\bq}}{{\bf q}}
\newcommand{{\tr}}{{\rm tr}}
\newcommand{{\evh}}{{E_{\rm vH}}}
\newcommand{\pprl}{Phys. Rev. Lett. \ }
\newcommand{\pprb}{Phys. Rev. {B}}

%\title{Electronic nematic order in FeSe}
\title{Interatomic Coulomb interaction and electron nematic bond order in FeSe}

\author{Kun Jiang}
\affiliation{Department of Physics, Boston College, Chestnut Hill, MA 02467, USA}
\author{Jiangping Hu}
\affiliation{
Institute of Physics, Chinese Academy of Sciences, Beijing 100190, China}
\author{Hong Ding}
\affiliation{
Institute of Physics, Chinese Academy of Sciences, Beijing 100190, China}
\author{Ziqiang Wang}
\affiliation{Department of Physics, Boston College, Chestnut Hill, MA 02467, USA}
\date{\today}

\begin{abstract}
Despite having the simplest atomic structure, bulk FeSe has an observed electronic structure with the largest deviation from the band theory predictions among all Fe-based superconductors and exhibits a low temperature nematic electronic state without intervening magnetic order. We show that the Fe-Fe interatomic Coulomb repulsion $V$ offers a natural explanation for the puzzling electron correlation effects in FeSe superconductors. It produces a strongly renormalized low-energy band structure where the van Hove singularity sits remarkably close to Fermi level in the high-temperature electron liquid phase as observed experimentally. This proximity enables the quantum fluctuations in $V$ to induce a rotational symmetry breaking electronic bond order in the $d$-wave channel. We argue that this emergent low-temperature $d$-wave bond nematic state, different from the commonly discussed ferro-orbital order and spin-nematicity, has been observed recently by several angle resolved photoemission experiments detecting the lifting of the band degeneracies at high symmetry points in the Brillouin zone. We present a symmetry analysis of the space group and identify the hidden antiunitary $T$-symmetry that protects the band degeneracy and the electronic order/interaction that can break the symmetry and lift the degeneracy. We show that the $d$-wave nematic bond order, together with the spin-orbit coupling, provide a unique explanation of the temperature dependence, momentum space anisotropy, and domain effects observed experimentally. We discuss the implications of our findings on the structural transition, the absence of magnetic order, and the intricate competition between nematicity and superconductivity in FeSe superconductors.

\typeout{polish abstract}
\end{abstract}


\maketitle

\section{Introduction}

The electron nematic phase with purely rotational symmetry breaking is arguably the most unconventional and poorly understood phase in Fe-based superconductors. In the Fe-pnictides, the direct observation of nematic electronic structure has been difficult since the orthorhombic lattice distortion is immediately followed by the collinear spin density wave (SDW) order that breaks, in addition to the spatial-orbital rotational symmetry, lattice translation, spin-rotation, and time-reversal symmetries. This leaves the origin of nematicity highly debated \cite{fernandes} between the spin-nematic \cite{hu08,matsuda} and ferro-orbital order \cite{brink,phillips,ku,kontani,devereaux} scenarios. In contrast, the Fe-chalcogenide FeSe undergoes the tetragonal to orthorhombic structural transition at $T_s=87 K$ and the SC transition at $T_c=9 K$ without any trace of magnetic order \cite{mcqueen,bohmer13}. The latter enabled direct observations of a rotational symmetry breaking electronic structure at low temperatures by several angle resolved photoemission spectroscopy (ARPES) experiments recently \cite{nakayama,shimojima,watson,ding,suzuki}, unveiling that the SC transition in bulk FeSe takes place from a highly unconventional nematic electronic state. Understanding the microscopic origin of this nematic order, its relation to the structural transition and magnetism is the focus of this work.

In a nutshell, ARPES detects the splitting of symmetry protected degeneracies between the $d_{xz}$ and $d_{yz}$ orbitals in the band dispersions at the high symmetry points M$(\pi,0,0)$ and the $\Gamma(0,0,0)/$Z$(0,0,\pi)$ in the original Brillouin zone (BZ). More importantly, the corresponding degeneracy splitting energy $\Delta_{M}$ and $\Delta_{\Gamma/Z}$ appear {\em anisotropic} in momentum space with different temperature dependence. The nematic transition is determined by the strongly $T$-dependent $\Delta_M$ whose onset coincides with \cite{shimojima,watson} or is about $20$K above the structural transition $T_s$ \cite{nakayama,ding}. $\Delta_M$ rises with decreasing $T$ and reaches $\Delta_M\simeq62$meV at $22$K, closely resembling the $T$-dependence of an energy scale associated with a symmetry breaking order parameter. This is consistent with NMR \cite{baek,bohmer15}, optics, and transport measurements \cite{wu,huynh} detecting changes in the electronic state near or above $T_s$. The splitting at the BZ center, $\Delta_{\Gamma}\simeq30$meV at $22$K, on the other hand, does not break rotational symmetry above $T_s$. It is nearly $T$-independent up to $150$K and insensitive to $T_s$ and the onset of $\Delta_M$ \cite{ding}. It was thus conjectured \cite{ding} that the nematic order in FeSe is not due to the commonly discussed ferro-orbital order \cite{baek,vafek14,andersen15}, but rather driven by a $d$-wave nematic bond order \cite{li} $O_{\rm dNB}=\sum_k (\cos k_x-\cos k_y)[n_{xz}(k)+n_{yz}(k)]$.
   \begin{figure}
      \begin{center}
    \fig{3.4in}{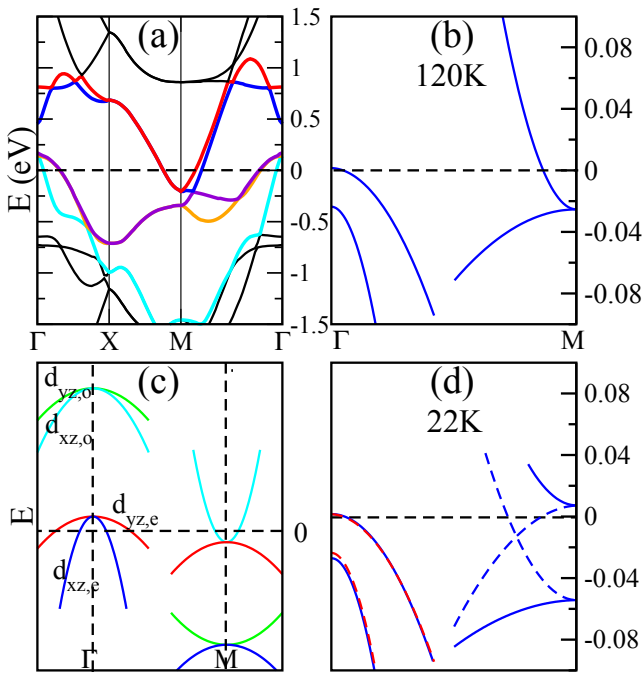}\caption{(a) LDA band structure of FeSe.(b) Band dispersions observed by ARPES in the symmetric phase at $120$K. (c) Band degeneracy at high symmetry points ($\Gamma, M$) among $d_{xz,e}$ (blue), $d_{yz,e}$ (red), $d_{xz,o}$ (cyan), and $d_{yz,o}$ (green).  (d) ARPES results in the nematic state at $22$K. Red dashed lines near $\Gamma$ are data taken at $120$K, showing $\Delta_{\Gamma}$ is nearly $T$-independent. Solid and dashed blue lines near $M$ correspond to the two domains. \label{fig1}}
    \end{center}
%    \vskip-0.5cm
    \end{figure}

We show in this work that the nearest neighbor Fe-Fe interatomic Coulomb repulsion $V$ can be the microscopic origin for the emergent nematic order, the absence of magnetism, and at a more fundamental level, the unusually large band renormalization in bulk FeSe. The observed electronic structure of FeSe shows the largest deviation from the local density approximation (LDA) band dispersions among all Fe-based superconductors \cite{eschrig,singh}. A remarkable difference from the Fe-pnictides is that the renormalization is the strongest at low energies near the Fermi level ($E_F$) as shown in Fig.~1, where the LDA bands are compared with ARPES measurements. Of crucial importance is the symmetry protected van Hove singularity (vHS) at $M$ point created by the saddle point in the band dispersions of the $d_{xz/yz}$ orbital. Being more than $250$meV below $E_F$ in LDA, it moves to a mere $\evh\simeq25$meV below the Fermi level in ARPES in the high temperature electron liquid phase at $T>T_s$. We show that this correlation-induced proximity of the vHS to $E_F$, concomitant with the large mass enhancement and the contraction of the FS pockets, sets the stage for the electronic nematic transition near $T_s$ that would ultimately gap out the vHS and induce the structural transition.

What is the microscopic interaction that would drive the vHS so close to $E_F$ in the high-temperature electron liquid phase? The more than $10$ times reduction in the distance from the hole band top at $\Gamma$ to the vHS at $M$ (see Fig.1) cannot come from crystal field corrections induced by local interactions \cite{sen05,sen10} since it comes from the same atomic orbital. Recent LDA+DMFT (dynamical meanfield theory) calculations \cite{yin11,yin12} show that the intra-atomic Hubbard $U$ and Hund's rule coupling $J$ would only produce a bandwidth reduction (albeit with some orbital dependence) and a mass enhancement by a factor of $3\sim4$ and leave the FS pockets much larger than those observed by ARPES and quantum oscillations \cite{watson}. Thus, intra-atomic correlations alone cannot account for the electronic structure near $E_F$ \cite{andersen15}.

We find that the nearest neighbor Coulomb $V$ generates directly hopping corrections to the band dispersion: it pushes the vHS at $M$ up toward $E_F$ and pulls down the top of the hole band at $\Gamma$, resulting in a low energy dispersion consistent with experiments. Both the FS pockets and the $\evh$ in ARPES can be produced by a $V\simeq0.73$eV in the Hartree-Fock (HF) theory with the bare LDA bandwidth $W\simeq4.2$eV. When the vHS is driven to $E_F$ (i.e. $\evh=0$), a nematic instability of the Pomeranchuk-type occurs in the symmetry breaking valence bond channels. The leading instability corresponds precisely to the $d$-wave nematic bond order. The proximity to this nematic instability in bulk FeSe allows us to carry out a weak-coupling analysis of the extended $t$-$U$-$V$ Hubbard model and find good agreement with experiments. Moreover, we find that the $V$-renormalized low-energy band structure with the reduced FS pockets promotes the $d$-wave nematicity while suppressing the collinear SDW, which is a possible explanation for the absence of magnetic order in bulk FeSe.

Since the symmetry content of the band degeneracies and its relation to the nematic order in Fe-based superconductors have not been understood, we begin in Section II with a systematic symmetry analysis. We show that the band degeneracy at the high symmetry point originates from the existence of two-dimensional irreducible representations of the space group involving rotation, reflection, and glide symmetries \cite{vafek}. We then show that there exists two sets of ``hidden'' antiunitary $T$-symmetries (borrowing the $T$ from an analogy to the time-reversal symmetry) that protect the band degeneracies at the $\Gamma$ and $M$ points respectively. All possible degeneracy lifting interactions are then studied according to their symmetry and symmetry-breaking properties in connection to the experimental findings on the temperature dependence, momentum space anisotropy, and domain effects. The analysis shows that the only interactions consistent with the experimental findings are the atomic spin-orbit coupling governing the lifting of the degeneracy at $\Gamma$ without breaking the four-fold rotation symmetry, and the rotational symmetry breaking $d$-wave nematic bond order that splits the degeneracy at $M$. In Section III, we develop the microscopic theory for the nematic state in bulk FeSe based on the $t$-$U$-$V$ Hubbard model and show that the quantum fluctuations in the intersite correlation $V$ leads to the important band renormalization and in particular to the dynamical $V$-driven proximity of the vHS to the Fermi level. The nematic instability is studied in detail and the obtained low energy band structure and the Fermi surfaces are compared to recent experimental results. In Section IV, we provide a summary and discuss the implications of these findings on FeSe films, the effects of electron doping, and the interplay between nematicity and superconductivity, as well as propose experimental tests for the theory.

\section{Band degeneracy, symmetry content, and effective interactions}

We begin with a discussion on the symmetry-protected band degeneracies at the $\Gamma$ and $M$ points illustrated in Fig.1(c). These symmetry properties dictate a rich and interesting set of possible degeneracy lifting interactions with different implications on the momentum space anisotropy and the domain effects. We will show that only the $d$-wave nematic bond order is naturally consistent with the experimental findings for the nematic transition.

The LDA electronic structure (Fig.1a) can be described by the tight-binding (TB) model down-folded to the Fe $3d$-manifold \cite{sen11,hu14},
\begin{equation}
H_t = \sum_{\sigma ij\alpha \beta } t_{ij}^{\alpha \beta}d_{i\alpha \sigma }^ \dagger d_{j\beta \sigma } + \sum_{\sigma i \alpha }\varepsilon _\alpha d_{i\alpha \sigma }^ \dagger d_{i\alpha \sigma },
\label{ht}
\end{equation}
where $\varepsilon _\alpha$ is the on-site (crystal field) energy of an electron in orbital $\alpha=(1,2,3,4,5)\equiv(xz,yz,{x^2}-{y^2},xy,{z^2})$ and
$t_{ij}^{\alpha \beta}$ is the electron hopping integral between sites $(i,j)$ and orbitals $(\alpha,\beta)$. The FeSe lattice structure contains two Fe atoms per unit cell labeled by $\ell=A$ and $B$. As a result, the TB Hamiltonian in momentum space is,
\begin{equation}
{H_t} = \sum_{k\sigma } {\psi _\sigma ^ \dagger (k){H_t}(k)} {\psi _\sigma }(k),
\label{htk}
\end{equation}
where the basis vector ${\psi _\sigma }(k) = [d_\sigma^A(k),d_\sigma^B(k)]^T $ with $d_\sigma^\ell =(d_{1\sigma}^\ell,d_{2\sigma}^\ell,\dots, d_{5\sigma}^\ell)$, and $H_t(k)$ is a $10\times10$ matrix whose eigenvalues govern the dispersion of the $10$ LDA bands,
\begin{equation}
H_t(k)\vert n,k\rangle=E_k^n\vert n,k\rangle, \qquad n=1\dots 10.
\label{enk}
\end{equation}
The explicit form of $H_t(k)$ and its parameters were derived in Refs.\cite{eschrig,hu14} for FeSe. Note that to avoid confusion, we use the $k$ values confined to the reduced two-Fe BZ shown in Fig.2(b) to label the momentum eigenstates such that the four $M$ points are located at $(\pm\pi,0)$ and $(0,\pm\pi)$ of the original one-Fe BZ.

\subsection{Symmetry protected band degeneracies}

The band degeneracies at $\Gamma$ and $M$ are related to three important space group symmetries $R\equiv S_4, \Sigma_d, G_s$ of the atomic and electronic structure in Fe-based superconductors. The most commonly discussed is the $S_4=\Sigma_z\cdot C_4$ where a four-fold rotation $C_4$ is followed by a mirror reflection about the $x$-$y$ plane to account for the staggering positions of the Se/As ions above and below the Fe-plane. The spatial-orbital operations of $S_4$ is therefore
\begin{eqnarray}
S_4 : \quad\quad x &\to& y,y \to  - x,z \to  - z, \nonumber \\
d^\ell_{xz}&\to& -d^\ell_{yz},d^\ell_{yz}\to d^\ell_{xz},d^\ell_{xy}\to-d^\ell_{xy}, \label{s4} \\
d^\ell_{x^2-y^2}&\to&-d^\ell_{x^2-y^2}, d^\ell_{z^2}\to d^\ell_{z^2}.
\nonumber
\end{eqnarray}
The less discussed point group symmetry $\Sigma_d$ is a mirror reflection about the diagonal $xy$-$z$ plane and operates in the spatial and orbital space according to,
\begin{eqnarray}
\Sigma_d: \quad\quad  x &\to& y,y \to x,z \to z, \nonumber \\
d^\ell_{xz}&\to& d^\ell_{yz}, d^\ell_{yz}\to d^\ell_{xz},d^\ell_{xy}\to d^\ell_{xy}, \label{sigmad} \\
d^\ell_{x^2-y^2}&\to&-d^\ell_{x^2-y^2}, d^\ell_{z^2}\to d^\ell_{z^2}.
\nonumber
\end{eqnarray}
We will show that $\Sigma_d$ plays as important a role as $S_4$ in the origin of the band degeneracy. Finally, the glide symmetry \cite{lee} of the space group $G_s=\Sigma_z \cdot T_{x,y}$ is associated with the translation by one lattice spacing along the $x$ or $y$ direction ($T_{x,y}$) followed by a mirror reflection about the $x$-$y$ plane. Under the $G_s=\Sigma_z\cdot T_x$ operation, $A\leftrightarrow B$ and
\begin{eqnarray}
G_s : \quad\quad x&\to& x+a, y\to y, z\to -z \nonumber \\
d^A_{xz}&\to&-d^B_{xz},d^A_{yz}\to-d^B_{yz},d^A_{xy}\to d^B_{xy}, \label{gs} \\
d^A_{x^2-y^2}&\to&d^B_{x^2-y^2}, d^A_{z^2}\to d^B_{z^2}.
\nonumber
\end{eqnarray}

Since $R$ is a symmetry of $H_t(k)$, $[H_t(k),R]=0$. Thus, $R\vert n,k\rangle=\vert n^\prime,k^\prime\rangle$ is a simultaneous eigenstate of $\vert n,k\rangle$ at the same energy $E_k^n$ as given in Eq.(\ref{enk}). Note that in general $k^\prime=Rk\ne k$. In particular, using Eqs.~(\ref{s4},\ref{sigmad},\ref{gs}), it is simple to show that in momentum space:
\begin{eqnarray}
S_4: k_x &\to& -k_y, k_y\to k_x \nonumber \\
\Sigma_d: k_x &\to& k_y, k_y\to k_x \label{kkprim}\\
G_s: k_x & \to & k_x, k_y\to k_y \nonumber
\end{eqnarray}
However, at certain high symmetry points in the BZ (e.g. $\Gamma$ and $M$), it is possible to have $k^\prime$ equal or equivalent to $k$ under the reciprocal lattice vector, resulting in symmetry related band degeneracies. To elucidate this, we transform the basis vector in Eq.~(\ref{htk}) into the eigenbasis of the glide symmetry operator $G_s$ via a unitary rotation $\Psi_\sigma(k)=U\psi_\sigma(k)=U[d_\sigma^A(k),d_\sigma^B(k)]^T$ \cite{lee}, where $U = {1\over\sqrt{2}}\left( {\begin{array}{*{20}{c}}
{{U_1}}&{{U_2}}\\
{{U_1}}&{ - {U_2}}
\end{array}} \right)$ and
\begin{equation}
{U_1} = \left( {\begin{array}{*{20}{c}}
1&0&0&0&0\\
0&1&0&0&0\\
0&0&1&0&0\\
0&0&0&1&0\\
0&0&0&0&1
\end{array}} \right),
{U_2} =\left( {\begin{array}{*{20}{c}}
1&0&0&0&0\\
0&1&0&0&0\\
0&0&{ - 1}&0&0\\
0&0&0&{ - 1}&0\\
0&0&0&0&{ - 1}
\end{array}} \right).
\nonumber
\end{equation}
The rotation mixes the orbitals defined on the $A$ and $B$ sublattices and turns them into the corresponding even-odd combinations, $d_{\alpha}^{e/o}= {1\over\sqrt{2}}(d_{\alpha}^A \pm d_{\alpha}^B)$, that form the eigenstates of $G_s$ with $\mp1$ eigenvalues: ${\Psi_\sigma }(k) = [d _{\sigma}^-(k),d_{\sigma}^+(k)]^T $ where
\begin{equation}
d_{\sigma}^{-/+}= (d_{xz\sigma}^{e/o},d_{yz\sigma}^{e/o},d_{x^2-y^2\sigma}^{o/e}, d_{xy\sigma}^{o/e},d_{z^2\sigma }^{o/e}).
\label{evenodd}
\end{equation}
Since $[H_t(k),G_s]=0$, $U$ block-diagonalizes $H_t(k)$ in the $\Psi_\sigma(k)$ basis,
\begin{equation}
H_k=U{H_t}(k){U^ \dagger } = \left( {\begin{array}{*{20}{c}}
{P(k)}&0\\
0&{P(k + Q)}
\end{array}} \right),
\label{uhu}
\end{equation}
where $Q=(\pi,\pi)$ is the reciprocal lattice vector for the two-Fe unit cell. Eqs. (\ref{evenodd}) and (\ref{uhu}) show that the lattice translation symmetry corresponds to their invariance under $k\to k+Q$, provided the exchange $e\leftrightarrow o$ is executed in orbital space \cite{young}. This results in the important identification $\vert d_{\alpha\sigma}^e,M^\prime\rangle=\vert d_{\alpha\sigma}^o,M^\prime+Q\rangle=\vert d_{\alpha\sigma}^o,M\rangle$ in the two-Fe zone. Diagonalizing $P(K)$ gives rise to five band dispersions with $-1$ eigenvalue under glide operation, and shifting them by $k\to k+Q$ generates the other five with $+1$ eigenvalue under $G_s$. These $10$-bands are shown in Fig.2(a) for FeSe by the solid and the dashed lines respectively. Note that the band degeneracies at $\Gamma$ are between the eigenstates within $P(k)$ (odd under $G_s$) or within $P(k+Q)$ (even under $G_s$), whereas the degeneracies at $M$ are between the eigenstates of $P(k)$ and $P(k+Q)$.

\begin{figure}
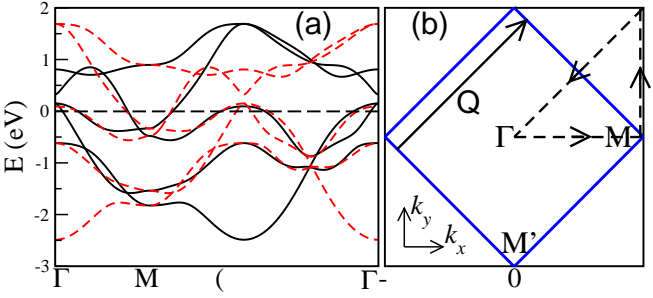

      \begin{center}
    \fig{3.4in}{fig2.eps}\caption{(a) TB band of FeSe. Black solid lines are from $P(k)$ and red dashed lines are from $P(k+Q)$. (b) BZ of FeSe. The larger zone bounded by the black lines is the one-Fe BZ while the smaller zone with blue boundaries is the two-Fe BZ. $Q$ corresponds to one of reciprocal lattice vectors in the two-Fe BZ. The dashed black lines with arrows indicate the trajectory over which the band dispersions are plotted in (a). \label{fig2}}
    \end{center}
%    \vskip-0.5cm
\end{figure}

On general grounds, the {\em two-fold} band degeneracies indicate that the space group of a Hamiltonian $H$ must have at least one two-dimensional irreducible representation. Specifically, there must exist at least two group generators, representable by two matrices $D_1$ and $D_2$ with $[H,D_1]=[H,D_2]=0$, that are mutually noncommuting, $[D_1,D_2]\neq0$. For the noninteracting system described by the TB Hamiltonian $H_t$, these two symmetries are manifestly the $S_4$ and $\Sigma_d$. When interactions are included, the degeneracies will remain so long as a pair of $D_{1,2}$ exist and the two-dimensional irreducible representation remains intact. For symmetry-breaking interactions that reduce the space group down to an abelian group with only one-dimensional representations, the band degeneracies will be lifted as we will show below.

Since the Hamiltonian $H_k$ in Eq.~(\ref{uhu}) is manifestly diagonal and the states in Eq.~(\ref{evenodd}) are the eigenstates at $\Gamma$ and $M$ points \cite{vafek}, the interplay between symmetry and band degeneracy can be studied by focusing on the $d_{xz}^{e/o}$ and $d_{yz}^{e/o}$ orbitals at $\Gamma$ and $M$ points separately. Eq.~(\ref{evenodd}) shows that the degeneracies at $\Gamma$ are within the e-e and o-o pairs, i.e. $\vert d_{xz}^e/d_{yz}^e,\Gamma\rangle$ and $\vert d_{xz}^o/d_{yz}^o,\Gamma\rangle$, whereas those at the $M$ point, with the $M^\prime$ folded to $M$ by the reciprocal lattice operation discussed above, are between the e-o components $\vert d_{xz}^o/d_{yz}^e, M\rangle$ and $\vert d_{xz}^e/d_{yz}^o,{M}\rangle$ as shown in Fig.1(c). These degeneracies clearly originate from the $S_4$ and the $\Sigma_d$ symmetries defined in Eqs.(\ref{s4}) and (\ref{sigmad}) since,
\begin{eqnarray}
S_4\vert d_{xz}^e,\Gamma\rangle&=&-\vert d_{yz}^e,\Gamma\rangle, \quad S_4\vert d_{yz}^e,\Gamma\rangle=\vert d_{xz}^e,\Gamma\rangle, \nonumber \\
\Sigma_d\vert d_{xz}^e,\Gamma\rangle&=&\vert d_{yz}^e,\Gamma\rangle, \quad \Sigma_d\vert d_{yz}^e,\Gamma\rangle=\vert d_{xz}^e,\Gamma\rangle, \nonumber
\end{eqnarray}
and similarly for the odd components at $\Gamma$; and
\begin{eqnarray}
S_4\vert d_{xz}^e,M\rangle&=&-\vert d_{yz}^o,M\rangle, \quad S_4\vert d_{yz}^e,M\rangle=\vert d_{xz}^o,M\rangle. \nonumber \\
\Sigma_d\vert d_{xz}^e,M\rangle&=&\vert d_{yz}^o,M\rangle, \quad \Sigma_d\vert d_{yz}^e,M\rangle=\vert d_{xz}^o,M\rangle, \nonumber
\end{eqnarray}
and similarly when ``$e$'' and ``$o$'' are interchanged at $M$, which is equivalent to $M^\prime$ under the reciprocal lattice vector. More importantly, it is crucial that the two symmetries do not commute, i.e. $[S_4,\Sigma_d]\neq0$ such that they form the two-dimensional irreducible representation of the space group necessary for the two-fold degeneracy.

For a deeper understanding that will facilitate the classification of the possible degeneracy-lifting interactions, let's recall the eigenbasis of the glide symmetry $\Psi_\sigma$ and Eq.~(\ref{evenodd}). In the decoupled subspace of $xz$ and $yz$ orbitals,$\Psi_\sigma=(d_{xz\sigma}^{e},d_{xz\sigma}^{o},d_{yz\sigma}^{e}, d_{yz\sigma}^{o})^T$. It is convenient to regard this four-spinor as the direct product of two two-spinors spanning the orbital space ($xz/yz$) and the sublattice space ($e/o$). In the rest of the discussion, the spin indices are suppressed for notational simplicity unless otherwise noted. The TB Hamiltonian $H_k$ in Eq.~(\ref{uhu}) can thus be written down explicitly up to constants at $\Gamma$ and $M$ points,
\begin{equation}
H_\Gamma   = \lambda _\Gamma\sigma _0 \otimes\tau _z,\quad
H_M   = \lambda_M \sigma _z \otimes \tau _z,
\label{hgammam}
\end{equation}
where $\sigma_i$ and $\tau_i$ are Pauli matrices acting in the orbital and sublattice spaces, respectively; and $\lambda_\Gamma$ and $\lambda_M$ are half the energy separations between the pair of degenerate points at $\Gamma$ and $M$. In this representation, the symmetry operators are
\begin{eqnarray}
(S_4,\Sigma_d, G_s)_\Gamma&=&(-i\sigma_y\otimes \tau_0,\sigma_x\otimes \tau_0,\sigma_0\otimes \tau_z), \label{symmetrymatrix1} \\
(S_4,\Sigma_d, G_s)_M&=&(-i\sigma_y\otimes \tau_x,\sigma_x\otimes \tau_x,\sigma_0\otimes \tau_z).
\label{symmetrymatrix2}
\end{eqnarray}
They commute with $H_{\Gamma/M}$ correspondingly.

Next, we show that the band degeneracies at $\Gamma$ and $M$ are protected by nonunitary symmetries in very much the same way that the time-reversal symmetry protects the $Z_2$ topological insulators \cite{kane}. Note that the direct product of the Pauli matrices comprises the identity and $15$ generators. A given $H_0=H_{\Gamma/M}$ anticommutes with $8$ of the other $14$ operators and commutes with the remaining $6$. Remarkably, the $8$ operators organize into two sets of $4$ operators; each forms an independent Clifford algebra with $H_0$ and is protected by the hidden antiunitary symmetry $T_1$ or $T_2$ with $T_{1,2}^2=-\bf{1}$ respectively. We can therefore group the generators according to the following
\begin{equation}
\underbrace{H_0,}_{\begin{array}{cc} T_1: &{\rm even}\\ T_2:&{\rm even}\end{array}} \underbrace{H_1,\dots,H_4}_{\begin{array}{cc} &{\rm even}\\ &{\rm odd}\end{array}},\underbrace{H_5,\dots,H_8}_{\begin{array}{cc}&{\rm odd} \\ &{\rm even}\end{array}},\underbrace{H_{9},\dots,H_{14}}_{\begin{array}{cc} &{\rm odd}\\ &{\rm odd}\end{array}},
\label{H15}
\end{equation}
where $[H_0,H_{9,\dots,14}]=0$, and
\begin{eqnarray}
&&\{H_0,H_{1,\dots,4}\}=0,\quad T_1 H_{0,1,\dots,4}T_1^{-1}=H_{0,1,\dots,4},
\nonumber \\
&&\{H_0,H_{5,\dots,8}\}=0, \quad T_2 H_{0,5,\dots,8} T_2^{-1}=H_{0,5,\dots,8},
\nonumber \\
&&T_{1}H_{5,\dots,14}T_{1}^{-1}=-H_{5,\dots,14}, \label{clifford} \\
&&T_2 H_{1,\dots,4,9,\dots14}T_2^{-1}=-H_{1,\dots,4,9,\dots14}.
\nonumber
\end{eqnarray}
Thus, the two-fold degeneracy of the quantum states at $\Gamma$ and $M$ in $H_0$ is protected by at least one necessary nonunitary symmetry $T\in (T_1,T_2)$, analogous to Kramers doublet $(\vert\alpha\rangle, T\vert\alpha\rangle)$ protected by the global nonunitary time-reversal symmetry. Specifically, we find that for $H_\Gamma$,
\begin{equation}
T_{1}^{\Gamma}=i\sigma_yK\otimes\tau_0, \quad T_2^\Gamma=i \sigma _y K \otimes \tau _z,
\label{t12gamma}
\end{equation}
where $K$ is the complex conjugation operator. They are effective time-reversal operators in the orbital angular momentum channel embedded symmetrically and antisymmetrically into the sublattice space respectively. Operators that are even under $T_{1,2}^\Gamma$ are
\begin{eqnarray}
T_1^\Gamma-{\rm even}:&& H_\Gamma,\sigma _0 \otimes \tau _x, \sigma _x \otimes\tau _y, \sigma _y \otimes\tau _y, \sigma _z\otimes \tau _y
\nonumber \\
T_2^\Gamma-{\rm even}:&& H_\Gamma, \sigma _0 \otimes\tau _y, \sigma _x\otimes \tau _x, \sigma _y\otimes \tau _x, \sigma _z\otimes \tau _x.
\nonumber
\end{eqnarray}
Similarly for $H_M$ at $M$ point,
\begin{equation}
T_{1}^{M}=\sigma_x \otimes i\tau_y K, \quad T_2^M=i \sigma _y K \otimes \tau _x,
\label{t12m}
\end{equation}
and
\begin{eqnarray}
T_1^M-{\rm even}:&& H_M,\sigma _x \otimes\tau _0, \sigma _y\otimes \tau _0, \sigma _z\otimes \tau _x, \sigma _z\otimes \tau _y,
\nonumber \\
T_2^M-{\rm even}:&& H_M, \sigma _y \otimes \tau _z, \sigma _x \otimes\tau _z, \sigma _0 \otimes\tau _x, \sigma _0\otimes \tau _y.
\nonumber
\end{eqnarray}
The identification of the antiunitary $T$-symmetries that protect the band degeneracies in Fe-based superconductors is one of the main results of this work. The finding allows us to characterize the form of electronic order induced by the possible effective interactions with respect to their ability to lift the band degeneracy at $\Gamma$ and $M$ points. The experimental observation of the momentum space anisotropy, temperature dependence, and domain features associated with the degeneracy lifting can then be used to determine the important microscopic electronic interactions responsible for nematicity in FeSe.
\begin{widetext}

\begin{table}[htb]
\begin{tabular}{c|c|c|c|c}
  \hline
  \hline
  Interactions at $\Gamma$/$M$ & Real Space & Order & $\Delta_{\Gamma}$ & $\Delta_M$ \\ \hline
  $g_1\{(d_{xz,e}^ \dagger {d_{xz,e}} + d_{xz,o}^ \dagger{d_{xz,o}})-(yz)\}$  & $ {g_1 (d_{xz,i}^\dag  d_{xz,i}  - d_{yz,i}^\dag  d_{yz,i} )}$ & Ferro-orbital & $\emptyset$ & $\emptyset$ \\
  $g_2\{(d_{xz,e}^\dagger {d_{xz,e}} - d_{xz,o}^\dagger {d_{xz,o}}) - (yz)\}$ & $ {g_2 (d_{xz,i}^\dag  d_{xz,i + \delta }  - d_{yz,i}^\dag  d_{yz,i + \delta })}$ & $s$-wave bond nematic & $\emptyset$ & $0$ \\
  $g_3\{(d_{xz,e}^\dagger {d_{xz,e}} - d_{xz,o}^\dagger {d_{xz,o}}) + (yz)\}$ & $ {g_3 ( - 1)^{\delta _y } (d_{xz,i}^\dag  d_{xz,i + \delta }  + d_{yz,i}^\dag  d_{yz,i + \delta })}$ & $d$-wave bond nematic & $0$ & $\emptyset$ \\ \hline
  ${g_{4+}^\prime}(d_{xz,e}^\dagger {d_{yz,e}} + d_{xz,o}^\dagger {d_{yz,o}}) + h.c.$  & ${g_{4+}^\prime}d_{xz,i}^\dag  d_{yz,i}  + h.c.$ & Orbital polarization & $\emptyset$ & $0$ \\
  ${ig_{4+}^{\prime\prime} }(d_{xz,e}^\dagger {d_{yz,e}} + d_{xz,o}^\dagger {d_{yz,o}}) + h.c.$  & $ig_{4+}^{\prime\prime} d_{xz,i}^\dag d_{yz,i}+h.c.$ & Ferro $L_z$/Spin orbit coupling & $\emptyset$ & $0$ \\
  ${g_{4-}^\prime }(d_{xz,e}^\dagger {d_{yz,e}} - d_{xz,o}^\dagger {d_{yz,o}}) + h.c.$  &  $g_{4-}^{\prime}(d_{xz,i}^\dag d_{yz,i+\delta}+d_{yz,i}^\dag d_{xz,i+\delta})  + h.c.$ & inter-orbital hopping & $\emptyset$ & $0$ \\
  ${ig_{4-}^{\prime\prime} }(d_{xz,e}^\dagger {d_{yz,e}} - d_{xz,o}^\dagger {d_{yz,o}}) + h.c.$  & $ig_{4-}^{\prime\prime}(d_{xz,i}^\dag d_{yz,i+\delta}-d_{yz,i}^\dag d_{xz,i+\delta})+h.c.$  & Orbital current/Spin orbital flux & $\emptyset$ & $0$ \\ \hline
  ${g_{5+}^\prime }(d_{xz,e}^\dagger {d_{xz,o}} + d_{yz,e}^\dagger {d_{yz,o}})+ h.c.$   & $g_{5+}^\prime e^{iQr_i}(d_{xz,i}^\dag d_{xz,i}+d_{yz,i}^\dag d_{yz,i})$ & Charge/spin density wave & $0$ & $0$ \\
  ${ig_{5+}^{\prime\prime} }(d_{xz,e}^\dagger {d_{xz,o}} + d_{yz,e}^\dagger {d_{yz,o}})+ h.c.$  & $ig_{5+}^{\prime\prime}(d_{xz,i}^\dag d_{xz,i+\delta}+d_{yz,i}^\dag d_{yz,i+\delta})+h.c.$ & circulating current (flux) & $0$ & $0$ \\
  ${g_{5-}^\prime }(d_{xz,e}^\dagger {d_{xz,o}} - d_{yz,e}^\dagger {d_{yz,o}})+ h.c.$   & $g_{5-}^\prime e^{iQr_i}(d_{xz,i}^\dag d_{xz,i}-d_{yz,i}^\dag d_{yz,i})$ & AF orbital & $0$ & $0$ \\
  ${ig_{5-}^{\prime\prime} }(d_{xz,e}^\dagger {d_{xz,o}} - d_{yz,e}^\dagger {d_{yz,o}})+ h.c.$   & $ig_{5-}^{\prime\prime}(d_{xz,i}^\dag d_{xz,i+\delta}-d_{yz,i}^\dag d_{yz,i+\delta})+h.c.$ & Ferro orbital current & $0$ & $0$ \\ \hline
  ${g_{6+}^\prime }(d_{xz,e}^\dagger {d_{yz,o}} + d_{xz,o}^\dagger {d_{yz,e}}) + h.c.$  & ${g_{6+}^\prime e^{iQr_i}d_{xz,i}^\dag  d_{yz,i} + h.c.}$ & AF orbital polarization & $0$ & $\emptyset$ \\
  ${ig_{6+}^{\prime\prime} }(d_{xz,e}^\dagger {d_{yz,o}} + d_{xz,o}^\dagger {d_{yz,e}}) + h.c.$  & $ig_{6+}^{\prime\prime}e^{iQr_i}d_{xz,i}^\dag  d_{yz,i}+h.c.$ & AF $L_z$/ AF spin orbit coupling & $0$ & $\emptyset$ \\
  ${g_{6-}^\prime }(d_{xz,e}^\dagger {d_{yz,o}} - d_{xz,o}^\dagger {d_{yz,e}}) + h.c.$  & ${g_{6-}^\prime(-1)^{\delta _y } (d_{xz,i}^\dag  d_{yz,i + \delta }- d_{yz,i}^\dag d_{xz,i + \delta})}+h.c.$ & $d$-wave interorbital hopping & $0$ & $\emptyset$ \\
  ${ig_{6-}^{\prime\prime} }(d_{xz,e}^\dagger {d_{yz,o}} - d_{xz,o}^\dagger {d_{yz,e}}) + h.c.$  & ${ig_{6-}^{\prime\prime}(-1)^{\delta _y } (d_{xz,i}^\dag  d_{yz,i + \delta }+ d_{yz,i}^\dag d_{xz,i + \delta})}+h.c.$ & $d$-wave orbital current & $0$ & $\emptyset$ \\ \hline \hline
\end{tabular}
\caption{The six types of effective interactions in the eigenbasis of glide symmetry (1st column), their real space representations (2nd column), physical meanings (3rd column),  and whether they generate degeneracy splitting at $\Gamma$ and $M$ (4th and 5th column).}
\end{table}
\end{widetext}

\subsection{Degeneracy lifting interactions}

%Having understood the symmetry content and the symmetry protection, we next turn to study the effective interactions that can lift the degeneracies.
On physical grounds, two degenerate states $\psi_1$ and $\psi_2$ can be split by either level shift or quantum mixing (hybridization). Using the spinor notation, $\psi  = {\left( {{\psi _1},{\psi _2}} \right)^T}$, we can express the two types of interactions as $g_z {\psi ^ \dagger }{\sigma _z}\psi$ and $g_x {\psi ^ \dagger}{\sigma _x}\psi$, respectively. The outcome of the $\sigma_z$ interaction dependents on the sign of $g_z$, which determines the relative position of $\psi_1$ and $\psi_2$ upon splitting and hence corresponds to two possible domains. On the other hand, the $\sigma_x$ interaction, being off diagonal, leads to a hybridized spectrum symmetric in $\psi_1$ and $\psi_2$ independent of the sign of $g_x$, and hence produces no domain effect. Since ARPES experiments observed two-domain contributions for the splitting at $M$ below $T_s$ and a splitting with a single domain at $\Gamma/Z$ above $T_s$, they must originate from two distinct type of interactions, i.e. $\Delta_M$ due to orbital shifting and $\Delta_{\Gamma/Z}$ due to quantum mixing of the orbitals.

It is therefore possible to write down all the possible forms of the electronic order or effective interactions between the degenerate orbital states at $\Gamma$ and $M$. Since the experimental observations are generally consistent with a unit cell containing two-Fe atoms, we will focus on the interactions that do not break the lattice translation symmetry. We find six types of relevant interactions or electronic orders among the $d_{xz}$ and $d_{yz}$ orbitals in the eigenbasis of the glide symmetry in Eq.~(\ref{evenodd}). They are listed in the first column of Table I with the corresponding coupling constants $g_1,\dots,g_6$. The coupling constants $g_{4,5,6}$ are in general complex and are thus decomposed into real and imaginary parts by writing $g_{m\pm}=g_{m\pm}^\prime +i g_{m\pm}^{\prime\prime}$, $m=4,5,6$. In the second column of Table I, the spatial representations that produce the corresponding interactions at the $\Gamma$ and $M$ points are given. In cases where the interaction is nonlocal, we consider the nearest neighbors of site $i$ indicated by $i+\delta_{x,y}$ along the $x$ and $y$ directions and the lowest angular momentum representations. The physical meaning of the electronic interaction/order is given in the third column and whether it can lift the band degeneracy at $\Gamma$ and $M$ points is noted in the last two columns of Table I.

In Table II, the matrix structures of the interactions are given explicitly in terms of the product of Pauli matrices in the orbital and sublattice subspaces. They span the complete set of the $15$ generators. The columns in Table II show the properties of the corresponding interaction under the point group and glide symmetry operations given in Eqs~(\ref{symmetrymatrix1}) and (\ref{symmetrymatrix2}). Note that although these symmetry operators can have different forms at $\Gamma$ and $M$ in the latter equations due to the presence of the reciprocal lattice vector, the symmetry and symmetry-breaking patterns of the interactions $g_i$ are the same at $\Gamma$ and $M$, which is consistent with the fact that the space group operations are global. However, it is important to realize that these symmetry operations can have different commutation relations at $\Gamma$ and $M$, thus resulting in different group properties such as whether there exists a two-dimensional irreducible representation.
%(Note that due to reciprocal lattice vector, symmetry operators have different forms at $\Gamma$ and $M$, which also may have different commuting rules at different points and different group properties.)
Since the band degeneracy is protected by the newly identified antiunitary $T$-symmetry, it is straightforward to determine if the interaction can lift the degeneracy by breaking the $T_{1,2}^\Gamma$ and $T_{1,2}^M$ at $\Gamma$ and $M$ points separately. In the case where the degeneracy remains, the protecting $T$-symmetry is displayed in the parenthesis.

\begin{widetext}

\begin{table}[htb]
\begin{tabular}{c||c|c|c|c|c|c|c|c|c|c|c|c|c|c|c}
  \hline \hline
  $$ & $g_1$ & $g_2$ & $g_3$ & $g_{4+}^{\prime}$ & $g_{4+}^{\prime\prime}$ & $g_{4-}^{\prime}$ & $g_{4-}^{\prime\prime}$    &$g_{5+}^{\prime}$ &$g_{5+}^{\prime\prime}$ & $g_{5-}^{\prime}$ & $g_{5-}^{\prime\prime}$ & $g_{6+}^{\prime}$ & $g_{6+}^{\prime\prime}$ & $g_{6-}^{\prime}$ & $g_{6-}^{\prime\prime}$ \\
  $$ & $\sigma_z \tau_0$ & $\sigma_z \tau_z$ & $\sigma_0 \tau_z$ & $\sigma _x \tau _0$ & $\sigma _y \tau _0$ & $\sigma _x \tau _z$ & $\sigma _y \tau _z$    &$\sigma _0 \tau _x$ &$\sigma _0 \tau _y$ & $\sigma _z \tau _x$ & $\sigma _z \tau _y$ & $\sigma _x \tau _x$ & $\sigma _y \tau _x$ & $\sigma _y \tau _y$ & $\sigma _x \tau _y$ \\ \hline
  $S_4$                & $\times$ & $\times$ & $\times$ & $\times$ & $\surd$  & $\times$ & $\surd$  & $\surd$  & $\surd$  & $\times$ & $\times$ & $\times$ & $\surd$  & $\surd$ & $\times$ \\
  $\Sigma_d$           & $\times$ & $\times$ & $\times$ & $\surd$  & $\times$ & $\surd$  & $\times$ & $\surd$  & $\surd$  & $\times$ & $\times$ & $\surd$  & $\times$ & $\times$  & $\surd$ \\
  $G_s$                  & $\surd$  & $\surd$  & $\surd$  & $\surd$  & $\surd$  & $\surd$  & $\surd$  & $\times$ & $\times$ & $\times$ & $\times$ & $\times$ & $\times$ & $\times$ & $\times$ \\
  $G_s\cdot S_4$               & $\times$ & $\times$ & $\times$ & $\times$ & $\surd$  & $\times$ & $\surd$  & $\times$ & $\times$ & $\surd$  & $\surd$  & $\surd$  & $\times$ & $\times$  & $\surd$\\
  $G_s\cdot \Sigma_d$          & $\times$ & $\times$ & $\times$ & $\surd$  & $\times$ & $\surd$  & $\times$ & $\times$ & $\times$ & $\surd$  & $\surd$  & $\times$ & $\surd$  & $\surd$ & $\times$ \\ \hline
  $\Delta_\Gamma $     &$\emptyset$ & $\emptyset$ & $0(*)$ & $\emptyset$  & $\emptyset$ & $\emptyset$  & $\emptyset$ & $0(T_1^\Gamma)$ & $0(T_2^\Gamma)$ & $0(T_2^\Gamma)$  & $0(T_1^\Gamma)$  & $0(T_2^\Gamma)$ & $0(T_2^\Gamma)$ & $0(T_1^\Gamma)$    & $0(T_1^\Gamma)$  \\
  $\Delta_M $          &$\emptyset$ & $0(*)$ & $\emptyset$ & $0(T_1^M)$   & $0(T_1^M)$  & $0(T_2^M)$    & $0(T_2^M)$    & $0(T_2^M)$  & $0(T_2^M)$      & $0(T_1^M)$  & $0(T_1^M)$    & $\emptyset$ & $\emptyset$ & $\emptyset$ & $\emptyset$  \\
  \hline \hline
\end{tabular}
\caption{Symmetry properties of the effective interactions in the direct product space of orbital and sublattice. The laster two rows indicate whether the band degeneracy at $\Gamma$ and $M$ is lifted ($\emptyset$) or not ($0$). In the latter case, the corresponding protecting antiunitary $T$-symmetry is given in the parenthesis, or a $*$ is given to indicate that the interaction is identically zero at the corresponding high symmetry point.}
\end{table}
\end{widetext}

\subsubsection{$g_{1,2,3}$ interactions}

Interactions $g_{1,2,3}$ produce orbital shifts as can be seen from Table I and break {\em both} $S_4$ and $\Sigma_d$ symmetries while preserving $G_s$ as shown in Table II. They are thus nematic interactions. Furthermore, since they break all the $T$-symmetries, the degeneracies at $\Gamma$ and $M$ are not protected and the lifting of the degeneracy must come with domain effects.
%which is only observed at $M$ points.
Thus, $g_{1,2,3}$ are only suitable candidates for describing the observations at $M$ points where two domains are observed by ARPES at low temperatures.

(i) The real space expression of $g_1$ is given in Table I,
\begin{equation}
O_{\rm FO}=g_1\sum_{i\sigma} (d_{xz,i\sigma}^\dag  d_{xz,i\sigma}  - d_{yz,i\sigma}^\dag  d_{yz,i\sigma}),
\label{ofo}
\end{equation}
which coincides with the commonly discussed ferro-orbital order (FO) parameter. It is isotropic in momentum space and leads to $\Delta_\Gamma=\Delta_M\neq0$ {\em simultaneously} by breaking all the $T$-symmetry, and is thus incompatible with the the experimental findings in FeSe.

(ii) As shown in Table I, $g_2$
%\begin{equation}
%O_s=g_2\sum_{i\eta\sigma} (d_{xz,i\sigma}^\dag  d_{xz,i + \eta\sigma}  - %d_{yz,i\sigma}^\dag  d_{yz,i + \eta\sigma } ),
%\label{osr}
%\end{equation}
%where $\eta=\hat x,\hat y$.
is a bond operator between the nearest neighbors. In momentum space,
\begin{equation}
O_{\rm sNB}=g_2\sum_{k\sigma} \gamma_k(d_{xz,k\sigma}^\dag  d_{xz,k\sigma}  - d_{yz,k\sigma}^\dag  d_{yz,k\sigma} ),
\label{osk}
\end{equation}
where $\gamma_k= \cos k_x + \cos k_y $. This is clearly an extended $s$-wave nematic bond order parameter that breaks both $S_4$ and $\Sigma_d$ symmetry. The $s$-wave form factor ($\gamma_k$) vanishes at $M$ point, which makes it possible for a nonzero expectation of $O_{\rm sNB}$ to generate a $\Delta_\Gamma\neq0$ but $\Delta_M=0$. Although this splitting patten is consistent with ARPES at $T>T_s$, $O_{\rm sNB}$ cannot describe the high temperature isotropic phase since it would break the four-fold rotation symmetry of the Fermi surface at $\Gamma$ with accompanying domain effects; both were not observed experimentally.

(iii) In momentum space, the bond interaction $g_3$ in Table I reads
%has its lowest angular momentum representation in real space as
%\begin{equation}
%O_{N}=g_3\sum_{i\eta\sigma } ( - 1)^{\eta_y } [d_{xz,i\sigma}^\dag  d_{xz,i + \eta \sigma}  + (xz\to yz)]
%d_{yz,i\sigma}^\dag  d_{yz,i + \eta\sigma } ).
%\label{odr}
%\end{equation}
%Fourier transform into momentum space, we obtain
\begin{equation}
O_{\rm dNB}=g_3\sum_{k\sigma} \beta_k(d_{xz,k,\sigma}^\dagger d_{xz,k\sigma}+d_{yz,k\sigma}^\dagger d_{yz,k\sigma}),
\label{nem}
\end{equation}
where $\beta_k=\cos k_x-\cos k_y$. This corresponds precisely to the $d$-wave nematic bond interaction that describes the low-temperature nematic state. $\langle O_{\rm dNB}\rangle\neq0$ leads to $\Delta_M\neq0$, but $\Delta_\Gamma=0$ since its form factor $\beta_k$ vanishes at the zone center. Note that $O_{\rm dNB}$ drives an {\em in-phase} $d$-wave bond order between the $d_{xz}$ and $d_{yz}$ orbital, which should be contrasted to the {\em out-of-phase} symmetry-preserving $d$-wave bond between these orbitals already present in the hopping terms of the TB model \cite{footnote}. In the next section, we will show how $O_{\rm dNB}$ can generated by the intersite Coulomb interaction, resulting in a $d$-wave nematic state consistent with experimental observations at low temperatures.

\subsubsection{$g_{4,5,6}$ interactions}

The remaining $3$ types of interactions, $g_{4,5,6}$ in Table I, generate quantum mixing/hybrdization among the degenerate orbitals. Thus any resulting degeneracy splitting would have only a single domain. Table II shows that although there are $4$ interactions in each type with different space group symmetry properties, the degeneracy lifting pattern is the same within each type.
%with $\Delta_\Gamma\neq0,\Delta_M=0$ for $g_4$; $\Delta_\Gamma=0,\Delta_M\neq0$ for $g_6$, and $\Delta_\Gamma=\Delta_M=0$ for $g_5$.
%
%Interactions of the types of $g_{4}$ and $g_{6}$ break either $S_4$ or $\Sigma_d$ symmetry but {\em not both}. Moreover, $g_{5}$ and $g_6$ break the glide symmetry $G_s$. All interactions of the $g_5$ type leaves one of the $T$-symmetry intact at either
%The coupling constants $g_{4,5,6}$ are in general complex and can be decomposed as $g_{m\pm}=g_{m\pm}^\prime +i g_{m\pm}^{\prime\prime}$, $m=4,5,6$.
%Indeed, the constraints of the observed properties discussed above limit the suitable interactions to $g_{4+}$ for $\Delta_\Gamma$ and $g_3$ for $\Delta_M$.

(i) Interactions of the $g_4$ type break either $S_4$ or $\Sigma_d$ symmetry while keeping the glide symmetry $G_s$ as seen in Table II. Thus, it is still possible for the remaining group to contain at least one two-dimensional irreducible representation. Indeed, one of the antiunitary symmetry in $T_{1,2}^M$ remains and protects the band degeneracy at $M$, whereas all $T$-symmetry is broken at $\Gamma$ where the band degeneracy will be lifted as indicated in Table II. From Table I, the spin SU(2) invariant representation of $g_{4+}^\prime$ describes the orbital polarization due to an effective on-site crystal field correction,
\begin{equation}
O_{\rm OP}=g_{4 + }^\prime\sum_{i\sigma} (d_{xz,i\sigma}^\dag  d_{yz,i\sigma}+{\rm h.c.}).
\label{op}
\end{equation}
Similarly, that of $g_{4-}^\prime$ generates an inter-orbital hopping or an extended $s$-wave orbital polarization,
\begin{equation}
O_{\rm sOP}=g_{4 - }^\prime\sum_{k\sigma} \gamma_k(d_{xz,k\sigma}^\dag  d_{yz,k\sigma}+{\rm h.c.}).
\label{sop}
\end{equation}
Although both $O_{\rm OP}$ and $O_{\rm sOP}$ split the degeneracy at $\Gamma$, the fact that they both break $S_4$ and $G_s\cdot S_4$ symmetries makes them incompatible with the experimental observation where the splitting at $\Gamma$ in the high temperature phase maintains the four-fold rotation symmetry.

Surprisingly, the imaginary components $g_{4\pm}^{\prime\prime}$ preserves $S_4$ symmetry despite of $\Delta_\Gamma\neq0$, which offers an intriguing, and the only possible account of the observed properties in the high temperature isotropic phase: degeneracy splitting only at $\Gamma$ point, coexisting four-fold symmetric Fermi surfaces, and the absence of domain effects. The spin SU(2) invariant representation of $g_{4+}^{\prime\prime}$ is
\begin{equation}
O_{L_z}=ig_{4+}^{\prime\prime}\sum_{i\sigma} (d_{xz,i\sigma}^\dag  d_{yz,i\sigma}-d_{yz,i\sigma}^\dag  d_{xz,i\sigma}),
\label{olz}
\end{equation}
which corresponds to an orbital angular momentum $L_z$ order that breaks the time-reversal symmetry. Remarkably, there exists a time-reversal invariant but spin-SU(2) breaking representation
\begin{equation}
O_{\rm soc}=i{g_{4+}^{\prime\prime}}\sum_{i\sigma}\sigma(d_{xz,i\sigma}^ \dagger {d_{yz,i\sigma}} -d_{yz,i\sigma}^\dagger {d_{xz,i\sigma}}),
\label{g4}
\end{equation}
which has an identical form as the spin-orbit interaction in the $d_{xz/yz}$ sector. Such an interaction can come from either the intrinsic SOC or be generated effectively \cite{vafek14}. Since the experimentally observed $\Delta_{\Gamma/Z}$ degeneracy splitting is nearly $T$-independent up to the highest measured temperature of $150$K \cite{ding,suzuki}, we conclude that the latter originates from the intrinsic atomic SOC and estimate the strength of the SOC to be on the order of $30$meV. Noted that since the SOC involving all $d$-orbitals breaks the glide symmetry, the corresponding band crossings should be lifted except at $M$ and $A$ point in the BZ \cite{vafek,bradley}. We note that although not all such splittings have been detected at the present time, the hybridization between the $d_{xy}$ and $d_{xz/yz}$ orbitals near $\Gamma/Z$ has indeed been observed by ARPES experiments \cite{ding,suzuki}.

The analysis of the interaction $g_{4-}^{\prime\prime}$ in Table I can be made in the same spirit. The spin SU(2) invariant realization of this bond operator corresponds to interorbital circulating current order (or the orbital flux phase) that breaks time-reversal symmetry. Similar to $O_{\rm soc}$ in Eq.~(\ref{g4}), there is a time reversal invariant realization that corresponds to spin-dependent orbital current order, where electrons with opposite spin-component traverse the lattice and accumulate opposite signs of the flux.

(iii) Interactions of the type $g_5$ and $g_6$ are hybridizations between the even and odd orbitals and thus break the glide symmetry $G_s$. The $g_5$ interactions are diagonal in the orbitals. The real parts represent charge density wave or spin density wave order ($g_{5+}^\prime$) at wave-vector $Q$ and antiferro (AF) orbital density wave order ($g_{5-}^\prime$), while the imaginary parts are realizations of the orbital current (or flux) order ($g_{5+}^{\prime\prime}$) and a hybrid of ferro-orbital and orbital current order ($g_{5-}^{\prime\prime}$), as shown in Table I. Note that all individual interaction of the $g_5$ type leaves one of the $T$-symmetry intact at $\Gamma$ and $M$ points as seen in Table II, and thus do not lift the degeneracy at $\Gamma$ and $M$. However, it is interesting to note that the coexistence of pairs of interactions, e.g. $g_{5\pm}^\prime$ or $g_{5\pm}^{\prime\prime}$, would remove all the $T$-symmetry protections and lead to simultaneous nonzero degeneracy splitting energies at $\Gamma$ and $M$.

(iv) The interactions of the $g_6$ type have the same orbital content as those of the $g_4$, but scatter between the even and odd components in the sublattice space. As shown in Table I, they generate AF orbital polarization ($g_{6+}^\prime$), AF spin orbital coupling or AF orbital angular momentum $L_z$ ($g_{6+}^{\prime\prime}$), interorbital hopping in the $d$-wave channel ($g_{6-}^\prime$), and $d$-wave orbital current or flux ($g_{6-}^{\prime\prime}$). The degeneracy at $\Gamma$ is protected by one of the remaining $T$-symmetry $T_{1,2}^\Gamma$, while it is lifted at $M$ since all $g_6$ interactions are odd under $T_{1,2}^M$. It is important to note that although $\Delta_\Gamma=0$ and $\Delta_M\neq0$, $g_6$ cannot account for the experimental observations at low temperatures since $S_4$ or $G_s\cdot S_4$ remains a symmetry which implies that the Fermi surfaces maintain four-fold symmetry without nematic distortions.

Based on the systematic symmetry analysis in this section, we conclude that the distinct degeneracy lifting observed by ARPES experiments at $\Gamma$ and $M$ originate from two independent interactions: the spin-orbit interaction in Eq.~(\ref{g4}) responsible for $\Delta_\Gamma\ne0$ already at high temperatures above $T_s$ and the $d$-wave bond nematic interaction in Eq.~(\ref{nem}) that produces the low-temperature nematic phase with $\Delta_M\neq0$.

\section{Intersite interaction and $d$-wave nematic bond order}

In the rest of the paper, we focus on how interatomic $V$ leads to strong band renormalization in FeSe and induces an emergent nematic order $\langle O_{\rm dNB}\rangle\neq0$ in the ground state. To this end, we study the extended Hubbard model
\begin{equation}
H=H_t+H_U+H_V,
\label{fullh}
\end{equation}
where $H_t$ is the tight-binding (TB) model given in Eq.~(\ref{ht}) and studied in the previous section. The intra-atomic interactions $H_U$ are given by the standard multi-orbital Hubbard model
\begin{eqnarray}
{H_U} &=& U\sum\limits_{i,\alpha }{{n_{i\alpha  \uparrow }}{n_{i\alpha  \downarrow }}}  + (U' - \frac{1}{2}J)\sum\limits_{i,\alpha  < \beta } {{n_{i\alpha }}{n_{i\beta }}}\label{hu} \\
 &-&J\sum_{i,\alpha\neq\beta}{\bf S}_{i\alpha}\cdot {\bf S}_{i\beta} + J\sum\limits_{i,\alpha  \ne \beta } {d_{i\alpha  \uparrow }^\dagger d_{i\alpha  \downarrow }^ \dagger {d_{i\beta  \downarrow }}{d_{i\beta  \uparrow }}}
\nonumber
\end{eqnarray}
where $U$ and $U^\prime$ are the on-site, intra- and inter-orbital on-site Coulomb repulsions and $J$ is the Hund's rule exchange coupling with $U = U' + 2J$. Note that when the Hamiltonian (\ref{hu}) is used to describe the complete set of $d$-orbitals, $J$ should be understood as an average of the exchange interactions of the $t_{2g}$ and the $e_g$ orbitals since the effects caused by the difference in the latter are usually small in a cubic system \cite{georges,coury}. The extended interatomic Coulomb interaction is given by
\begin{equation}
H_V = V\sum\limits_{ \langle i,j \rangle } :{{n_{i}}{n_{j}}}:
\label{hv}
\end{equation}
where the ``normal-order'' sign indicates that the direct Hartree term depending on the total density $n_i=\sum_\alpha n_{i\alpha}$ is subtracted, since that part of the interaction has been already included in the LDA. The same is true when treating $H_U$ in the HF theory \cite{sen05,sen10}. Thus, our treatment of interactions is in the same spirit as the LDA+U+V approach \cite{cococcioni,poteryaev}. The importance of the extended Coulomb interaction in Fe-based superconductors has been emphasized previously \cite{sen11} with a focus on the properties associated with such $p$-$d$ charge transfer metals. In the down-folded Fe-only model studied here, the interatomic interaction in Eq.~(\ref{hv}) is between the nearest-neighbor Fe atoms.

\subsection{Quantum fluctuations due to intersite $V$}

It is well known that the Fe-pinictides band structure is prone to a collinear SDW order that is also present in the multiorbital Hubbard model \cite{dagotto,sen10}. Writing $\langle {c_{i\alpha \sigma }^\dagger {c_{i\beta \sigma^\prime}}} \rangle  = {1\over 2}[n_\alpha+\sigma m_\alpha \cos{({Q}_{\rm AF}\cdot{\bf r}_i)}]\delta_{\alpha\beta}\delta_{\sigma\sigma'}$, where ${Q}_{\rm AF}=(\pi,0)$ and $n_\alpha$ and $m_\alpha$ are the density and spin density in orbital $\alpha$, a nonzero $m_\alpha$ is most easily obtained in the weak-coupling Hartree-Fock theory \cite{dagotto} which is reliable when $U$ is small. Since the electron correlation strength is comparable to the bandwidth of the $d$-electron complex, a complete description of the collinear SDW {\em metal} phase with realistic parameters would require a strong coupling approach that takes into account the correlation effects nonperturbatively \cite{sen10}. Here, we will carry out the Hartree-Fock theory for the extended Hubbard model in Eq.~(\ref{fullh}) using as effective parameters $U=1.4$eV and $J=0.2$eV. We will show that the quantum fluctuations induced by the inter-site $V$ lead to a renormalized band structure where the Fermi level sits close to the vHS, which allows a weak-coupling approach to capture the leading instability, the suppression of collinear SDW, and the emergence of the $d$-wave nematic bond order. The Hubbard interaction $H_U$ in Eq.~(\ref{hu}) is decoupled in terms of the self-consistent internal fields in the charge and spin sectors \cite{sen10}:
$\Delta_\alpha={1\over 2}(2U-5J)n -{1\over 2}(U-5J) n_\alpha$ and $h_\alpha={1\over 2}Jm+{1\over 2}(U-J) m_\alpha$, where $(n,m)=\sum_\alpha(n_\alpha,m_\alpha)$.
%We set $U=1.4$eV and $J=0.2$eV in this paper.
The self-consistently determined ground state indeed has ${Q}_{\rm AF}$-SDW order {\em in the absence} of the inter-site interaction $V$. To treat the quantum fluctuations beyond LDA, we decouple $H_V$ in Eq.~(\ref{hv}) in the hopping/bond channel,
\begin{equation}
H_V=-V\sum_{\langle i,j\rangle,\alpha\beta}(\chi_{ij}^{\alpha\beta} d_{i\alpha}^\dagger d_{j\beta}+h.c.-\vert\chi_{ij}^{\alpha\beta}\vert^2),
\label{vbond}
\end{equation}
where $\chi_{ij}^{\alpha\beta}=\langle d_{j\beta}^\dagger d_{i\alpha}\rangle$ and the spin index is suppressed for simplicity unless otherwise noted. In the presence of translation symmetry, the real nearest neighbor valence bond between any pair of orbitals $(\alpha,\beta)$ can be decomposed into the lattice harmonics of different angular momentum in $k$-space:
\begin{equation}
\chi_{ij}=\sum_k [2\chi_s \gamma_k+2 \chi_{px}\eta_{k}^x+2 \chi_{py}\eta_k^y+ 2\chi_d\beta_k],
\label{chi}
\end{equation}
where $\gamma_k$ and $\beta_k$ are the extended $s$-wave and $d$-wave form factors and $\eta_k^{x(y)}=i\sin k_{x(y)}$ are the $p$-wave form factors. Comparing Eqs ~(\ref{vbond}) and (\ref{chi}) to the tight-binding model in Eq.(\ref{ht}), it is clear that such quantum fluctuations amount to renormalizing the hopping integral
\begin{equation}
{\bf t}_{ij}^{\alpha\beta}=t_{ij}^{\alpha\beta}-V\chi_{ij}^{\alpha\beta}
\label{bt}
\end{equation}
between the nearest neighbors. There are in fact two classes of contributions generated by the inter-site interaction analogous to the situation in a general renormalization group analysis: (i) the corrections to the existing hopping parameters that maintain the lattice symmetry and thus lead to the renormalization of the band structure, and (ii) the spontaneous generation of new and symmetry breaking hopping channels. Correspondingly, we can write
$$
H_V=H_V^{\rm b.r.}+H_V^{\rm s.b.}.
$$
Since the full expressions for $H_V^{\rm b.r.}$ and $H_V^{\rm s.b.}$ are rather lengthy, we shall display explicitly the terms involving only the $t_{2g}$ orbitals:
\begin{widetext}
\begin{eqnarray}
H_V^{\rm b.r.}=&-&V\sum_k 2 \bigl[\chi_s\gamma_k(d_{xz,k}^\dagger d_{xz,k}+d_{yz,k}^\dagger d_{yz,k})
+\chi_d\beta_k(d_{xz,k}^\dagger d_{xz,k}-d_{yz,k}^\dagger d_{yz,k})
+\chi_{px}^{14}\eta_k^{x} d_{xz,k}^\dagger d_{xy,k} \nonumber \\
&+&\chi_{py}^{24}\eta_k^{y} d_{yz,k}^\dagger d_{xy,k}
+\chi_s^{44}\gamma_kd_{xy,k}^\dagger d_{xy,k}+ h.c.] + ({\rm terms\ involving\ }e_g\ {\rm orbitals}) \bigr],
\label{hvbr}
\end{eqnarray}
where $\chi_s=(\chi_s^{11}+\chi_s^{22})/2$ and $\chi_d=(\chi_d^{11}-\chi_d^{22})/2$. The symmetry-breaking part is
\begin{eqnarray}
%H_V^{\rm S.B.}=&-V_0&\sum_k\bigl[2\Delta_s\gamma_k (d_{xz,k}^\dagger d_{xz,k}-d_{yz,k}^\dagger d_{yz,k})\nonumber \\
%&+&2\Delta_d\beta_k(d_{xz,k}^\dagger d_{xz,k}+d_{yz,k}^\dagger d_{yz,k})\nonumber \\
%&+& 2\chi_d^{33} \beta_k d_{xy,k}^\dagger d_{xy,k} + \cdots],
%\label{hvsb}
H_V^{\rm s.b.}=&-&V_0\sum_k2\bigl[\Delta_s\gamma_k (d_{xz,k}^\dagger d_{xz,k}-d_{yz,k}^\dagger d_{yz,k})
+\Delta_d\beta_k(d_{xz,k}^\dagger d_{xz,k}+d_{yz,k}^\dagger d_{yz,k}) \nonumber \\
&+&(\chi_{px}^{11/22} \eta_k^{x} +\chi_{py}^{11/22} \eta_k^{y})d_{xz/yz,k}^\dagger d_{xz/yz,k}
+(\chi_s^{12} \gamma_k +\chi_d^{12}\beta_k+\chi_{px}^{12} \eta_k^{x}
+\chi_{py}^{12} \eta_k^{y})d_{xz,k}^\dagger d_{yz,k}\nonumber \\
&+&(\chi_{s}^{14/24} \gamma_k+\chi_{d}^{14/24} \beta_k) d_{xz/yz,k}^\dagger d_{xy,k}
+\chi_{py}^{14}\eta_k^{y} d_{xz,k}^\dagger d_{xy,k}+\chi_{px}^{24}\eta_k^{x} d_{yz,k}^\dagger d_{xy,k} \label{hvsb}\\
&+&(\chi_{px}^{44} \eta_k^{x} +\chi_{py}^{44} \eta_k^{y} )d_{xy,k}^\dagger d_{xy,k}
+\chi_d^{44} \beta_k d_{xy,k}^\dagger d_{xy,k} + h.c. + ({\rm terms\ involving\ }e_g\ {\rm orbitals}) \bigr],
\nonumber
\end{eqnarray}
\end{widetext}
where $\Delta_s=(\chi_s^{11}-\chi_s^{22})/2$ and $\Delta_d=(\chi_d^{11}+\chi_d^{22})/2$. The second term in Eq.(\ref{hvsb}) is precisely the $d$-wave nematic interaction in Eq.~(\ref{nem}) and the last term is its counterpart in the $d_{xy}$ channel. Note that we have denoted the interaction strength differently as $V$ and $V_0$ in Eqs~(\ref{hvbr}) and (\ref{hvsb}). Although the bare values stemming from the microscopic Coulomb interaction are expected to the same, the {\em effective} interaction strengths $V_0\neq V$ due to the effects of orbital polarization, screening, and other orbital dependent contributions. To develop more physical insights, we will vary $V$ and $V_0$ independently around $V_0/V=1$ in the calculations.

\subsubsection{Renormalization of the band structure in bulk FeSe}

We first discuss how $V$ renormalizes the band structure by switching off $H_V^{\rm s.b.}$ (i.e. setting $V_0=0$). The evolution of the self-consistently determined low-energy band dispersions are shown in Fig.~3. With increasing $V$, the value of $\evh\simeq350$meV in the TB model at $V=0$ shown in Fig.~3(a) is renormalized close to the experimental value of $25$meV at $V=0.73$eV in Fig.~3(b), and then to coincide with the Fermi level $E_F$ at $V=0.763$eV shown in Fig.~3(c). With increasing $V$, notice that the Dirac crossings of the $d_{xy}$ and $d_{yz/xz}$ bands also move up toward the Fermi level. Eventually, the vHS is pushed above the Fermi level which now cuts through the Dirac nodes, while the hole bands sink below $E_F$ at $V=0.85$eV as shown in Fig.~3(d), realizing an interesting state of a Dirac semimetal. Correspondingly, Figs.3(e)-(h) show the remarkable evolution of the four-fold symmetric Fermi surfaces: from the very large LDA hole and electron pockets in Fig.~3(e) to the observed small elliptical electron pockets with prominent quasi-1D character in Fig.~3(f), and then to four flower pedals at $\evh=0$ in Fig.~3(g), and finally to four Dirac points in Fig.~3(h).

\begin{figure}
      \begin{center}
    \fig{3.4in}{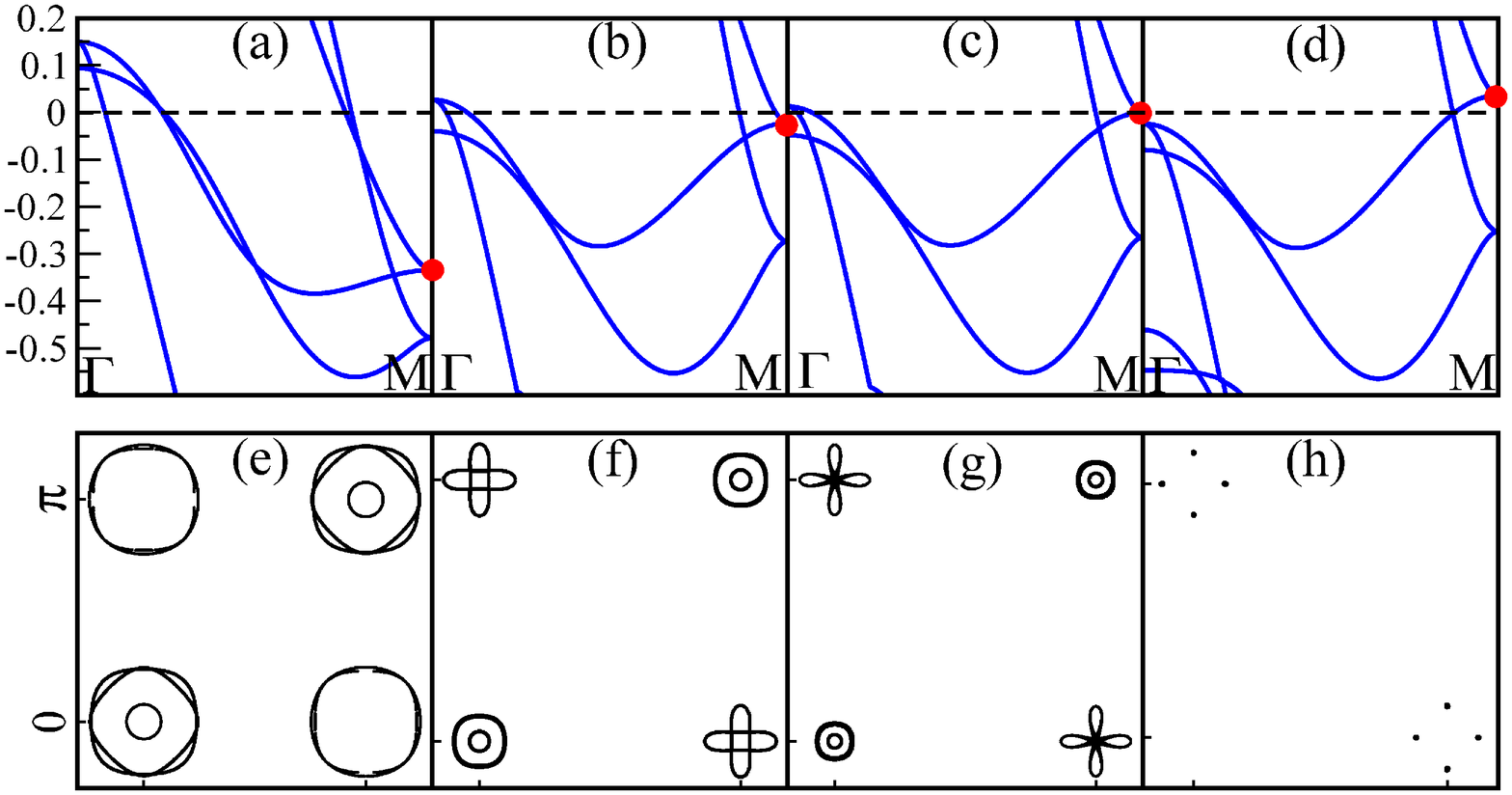}
    \caption{ Renormalized band structure (a-d) in unit of eV and corresponding FS (e-h) at $V=0$ (a), $0.73$eV (b), $0.763$eV (c), and $0.85$eV (d). Red dots indicate vHS at $d_{xz/yz}$ degeneracy point at $M$.
    FSs in (e) are $87.5$\% of their actual sizes. }
    \end{center}
%    \vskip-0.5cm
    \end{figure}

Thus, the immediate consequence of the strong band renormalization in bulk FeSe is the dynamically generated proximity of the vHS to $E_F$ in the electron liquid phase. It is remarkable that for the range of $V$ values where the vHS is within $25$meV of $E_F$ in the renormalized band structure, the self-consistent solution of the ground state is always an electron liquid state that is stable against collinear SDW and charge density wave order. In Fig.~4(a), the 2D density of states (DOS) of the renormalized band at $\evh=0$ is plotted, showing its logarithmic divergence at $E_F$. Correspondingly, the calculated two-particle static susceptibility $\chi_0({\bf q})$ shown in Fig.~4(b) displays a sharp peak at ${\bf q}=0$ due to the vHS, which appears even stronger than the peak at $(\pi,0)$ which dominated the DOS in the absence of $V$. Thus, the electron liquid state is only unstable toward a ${\bf q}=0$ nematic instability, analogous to the nematicity due to the vHS proposed for bilayer Sr-ruthenates \cite{kivelson}. Interestingly, we find that the vHS strengthens considerably when the hopping integral between the $d_{xz/yz}$ and $d_{xy}$ orbitals, described by the term $-2it_x^{14}\sin ({k_{x/y}})d_{xz/yz}^ + {d_{xy}}$ in the TB model, is decreased. The TB fit to LDA bands gives $t_x^{14}=305$meV. Reducing $t_x^{14}$ by a factor of $3$ leads to a significantly flatter $d_{xy}$ band near $\Gamma/Z$ as observed in the ARPES experiments \cite{ding} and to reduced curvatures of the $d_{xz/yz}$ band at the vHS point. Figs~4(a)-(b) show that the corresponding DOS and the ${\bf q}=0$ susceptibility are significantly enhanced, suggesting a much stronger nematic instability of the Pomeranchuk-type.

\subsubsection{$d$-wave Bond Nematic Order}

We next show that the ${\bf q}=0$ instability corresponds to the $d$-wave nematic bond order. To this end, we first set $V=0.763$eV where $\evh=0$ (Fig.3c) and switch on $V_0$ in the symmetry breaking part of the Hamiltonian $H_V^{\rm s.b.}$ in Eq.(\ref{hvsb}). The self-consistent solutions show that among all the symmetry breaking terms, the leading instability occurs precisely in the $d$-wave nematic bond channel with nonzero $\Delta_d$ and $\chi_d^{44}$ in Eq.(\ref{hvsb}), which has the largest form factor at $M$-point. The degeneracy splitting energy $\Delta_M$ between the $d_{xz}$ and $d_{yz}$ orbitals
%associated with the breaking of the $S_4$ symmetry
is calculated directly from the self-consistently determined eigenstate energies at $M$-point and plotted in Fig.~4(c) as a function of $V_0$. It serves as a quantitative measure of the degree of nematicity in the ground state.

To explore the range of $V$ over which the nematicity is controlled by the proximity of the vHS to $E_F$, we plot in Fig.~4(d) the $\Delta_M$-map as a function of $V$ for different ratios of $V_0/V$. The distance of the vHS to $E_F$, i.e. $\evh$, varies with $V$, and the dashed vertical line marks the location where the vHS sits at $E_F$, i.e. where $\evh=0$. The most notable feature of Fig.~4(d) is the existence of plateaux nearly symmetrically distributed around $\evh=0$ that grows in width and height with increasing $V_0/V$. For $V_0/V\sim1$, the latter covers the experimentally observed $\evh\simeq25$meV. Parallel results are obtained for $t_x^{14}=305$meV where a similar $\Delta_M$ requires a larger $V_0/V$.

\begin{figure}
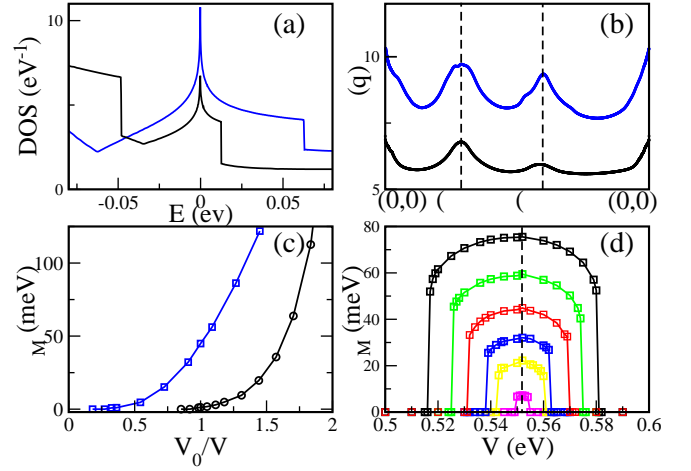

      \begin{center}
    \fig{3.4in}{fig4.eps}\caption{ DOS (a) and static susceptibility (b) of $V$-renormalized band structure at $\evh=0$ and corresponding degeneracy-splitting energy $\Delta_M$ (c). Black lines in (a)-(c) for TB fit $t_x^{14}=305$meV and blue line for $t_x^{14}=100$meV. (d) $\Delta_M$ versus $V$ around vHS for $t_x^{14}=100$meV; curves from bottom to top are for $V_0/V=0.6,0.8,0.9,1.0,1.1$, and $1.2$. The dashed black vertical line in (d) indicates where the vHS coincides with the Fermi level.}
    \end{center}
%   \vskip-0.5cm
    \end{figure}

In Fig.~5, we plot the band dispersion and the FS calculated self-consistently in the $d$-wave bond nematic ground state {\em in a single-domain}. In Figs.~5(a)-(b), the renormalized $E_F$ of the electron liquid is at the vHS, i.e. $\evh=0$. The $d$-wave nematic order removes the vHS by splitting the band-degeneracy around the Fermi level and produces the $4$-fold symmetry breaking FS pockets as a result of the $d$-wave Pomeranchuk distortion. In Figs.~5(c)-(d), $\evh=25$meV in the electron liquid phase, which corresponds to the observed value for the nominally undoped FeSe. Figs.~5(e)-(f) are obtained for $1\%$ electron doped Fe$_{1.01}$Se which may be closer to the as-grown samples used experimentally \cite{mcqueen,bohmer13,ding}. In the latter two cases, we find good agreements with the measured dispersions for a single domain shown in Fig.~1(d) near the $M$ point as well as the shape of the FS pockets \cite{watson,suzuki,ding}.

It is important to point out that the $d$-wave nematic bond order and the ferro-orbital order are {\em not} orthogonal and mutually exclusive. They belong to the same space group as it is clear from Table II that the interactions $g_{1,2,3}$ have the same symmetry breaking pattern. As a result, both $\Delta_d =\sum_{k}\beta_k [n_{xz}(k)+n_{yz}(k)]$ and $\Delta_{FO}=\sum_k[n_{xz}(k)-n_{yz}(k)]$ are none-zero in the $d$-wave bond nematic state, deriving from a momentum distribution function $n_\alpha(k)$ that breaks the $S_4$ symmetry, unless additional particle-hole symmetry is present which is not the case in FeSe. The conventional Pomeranchuk FS distortion is thus generalized to the case with orbital-lattice momentum coupling. This explains why local probes such as NMR have detected FO order in the low temperature nematic phase \cite{baek, bohmer15}. We stress that, in the present theory, the nematic transition is driven by the $d$-wave bond order that couples to the intersite Coulomb interaction and the FO order is induced parasitically, since the energy lowering in going from the electron liquid to the nematic state primarily comes from the $d$-wave nematic bond order. Indeed, since the FO order parameter $\Delta_{\rm FO}$ only couples directly to the local intra-atomic interactions but not to $V$, the degeneracy splitting energy $\Delta_M = 8V_0\Delta_d -{1\over2}(U-5J)\Delta_{\rm FO}$ in the HF theory is dominated by the contribution from $\Delta_d$, and even more so due to the large Hund's rule coupling $J$ in Fe-base superconductors \cite{sen10,haule} making $U-5J$ much smaller. This explains the insensitivity of degeneracy splitting energy $\Delta_\Gamma$ at the zone center to the $d$-wave nematic bond order at low temperatures, which is estimated to have only about $5$ meV variations from $120$K down to $22$K in the ARPES data \cite{ding}, as shown in Fig.~1(d). Nevertheless, the induced $\Delta_{\rm FO}$ does lead to the distortion of the hole FS near the zone center. For the values of $U$ and $J$ studied here, the hole FS pockets shown in Figs~5(d), 5(f), and 5(h) near the $\Gamma$ point indeed break the 4-fold rotational symmetry in a manner consistent with the experimental observations \cite{suzuki}.
\begin{figure}
      \begin{center}
    \fig{3.4in}{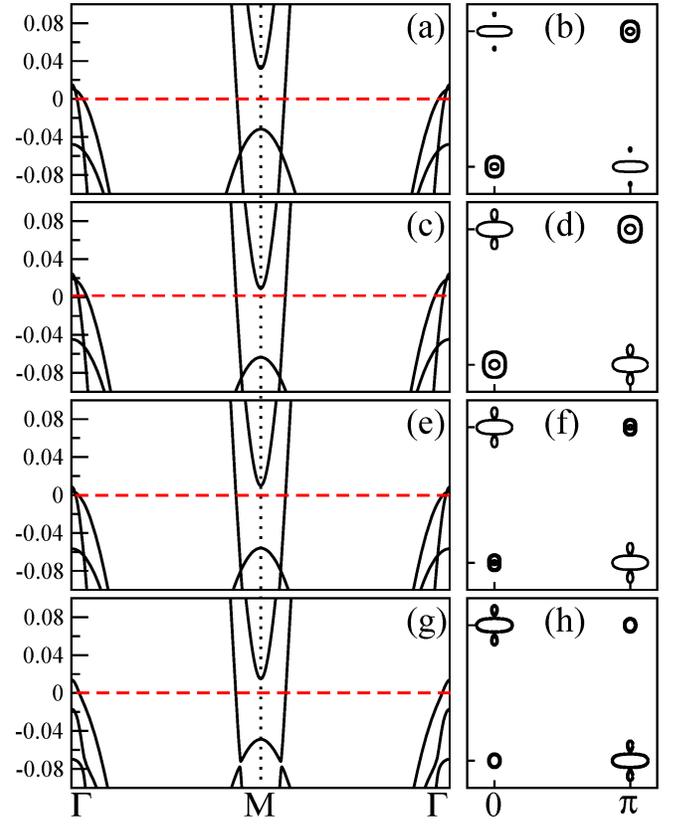}\caption{Band dispersion in unit of eV (left panel) and FS (right panel) in $d$-wave nematic state. (a) and (b): $V=0.763$eV, $\evh=0$, and $V_0=1.3$eV. (c) and (d): $V=0.73$eV, $\evh=25$meV, and $V_0=1.34$eV. (e) and (f): $V=0.75$ and $V_0=1.325$eV at 1\% electron doping. (g) and (h): With SOC $\lambda_{\rm soc}=28$meV, $V=0.763$eV, $V_0=1.32$eV at 1\% electron doping.}
    \end{center}
%    \vskip-0.5cm
    \end{figure}

In Figs 5(g)-(h), we show the results obtained when an atomic SOC term involving the entire 3$d$ complex, $H_{\rm soc}=
\sum\limits_{i\alpha \beta \sigma \sigma '} {\lambda _{soc} } \left\langle \alpha  \right|L\left| \beta  \right\rangle \left\langle \sigma  \right|S\left| {\sigma '} \right\rangle d_{i\alpha \sigma }^ +  d_{i\beta \sigma '}$, is added to the Hamiltonian with $\lambda_{\rm soc}\simeq28$meV. As predicted by the symmetry analysis in the last section, the SOC splits the band degeneracy at $\Gamma$ without affecting the the bands at $M$. Moreover, it pushes one of the two hole bands at $\Gamma$ below $E_F$, leaving a single two-fold symmetric hole pocket consistent with what was observed in ARPES and quantum oscillation experiments \cite{watson,suzuki}. It is important to note that since the glide symmetry is broken by the SOC, the Dirac crossings located below the Fermi level between the $d_{xy}$ and $d_{yz}$ orbitals in a single domain are lifted by the SOC, as shown in Fig.~5(g). This small gapping of the Dirac points can serve as a landmark for the presence of a sizable SOC in FeSe, although the detection of the $d_{xy}$ band by ARPES below the Fermi level has been notoriously difficult in {\em bulk} FeSe. In contrast, we expect that Dirac like crossings formed by the $d_{xy}$ and $d_{yz}$ bands originating from different domains will not be gapped.

Finally, for all the cases studied with $\evh$ within $25$meV of the Fermi level, the $d$-wave nematic order dominates and the collinear SDW order is absent in the self-consistent solutions of the ground state. It is thus highly conceivable that magnetism and $d$-wave nematicity are competing caricatures in FeSe superconductors. In the present theory, the strong band renormalization, the suppression of the collinear magnetic order, and the emergence of the electronic nematic order have the common origin which is the inter-site Coulomb interaction.

\section{Summary and Discussions}

We have shown that the rise and the demise of symmetry protected degeneracies in the electronic band structure can be used to probe the novel quantum states and the underlying interactions in correlated multiorbital electron materials. This work makes two advances in this direction with specific emphasis on Fe-based superconductors. First, a systematic symmetry analysis revealed the ``hidden'' antiunitary $T$-symmetries that protect the degeneracies at high-symmetry points in the BZ, and their connection to point-group and glide symmetry operations. This enabled the identification of the relevant electronic order/interaction that can break the $T$-symmetry and lift the band degeneracy. These results are applicable to all Fe-based superconductors. For bulk FeSe, the above analysis combined with recent experimental observation of the splitting of the band degeneracy, including their momentum space anisotropy, temperature dependence, and domain effects lead uniquely to the conclusion that the splitting present already in the high temperature electron liquid phase at $\Gamma$ is due to the atomic SOC, while the splitting at $M$ is due to the $d$-wave nematic bond order that emerges only in the low temperature nematic phase.
\begin{figure}
      \begin{center}
    \fig{3.4in}{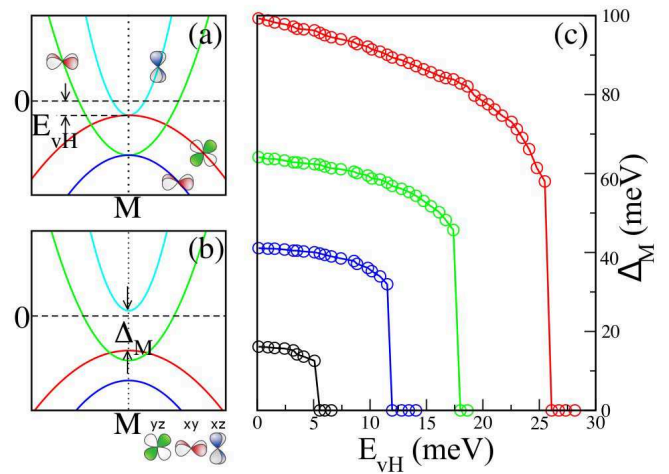}\caption{(a) Schematics of the electronic structure near $M$ showing the location of the vHS ($\evh$) before (a) and the degeneracy splitting energy $\Delta_M$ after (b) the nematic transition. (c) $\Delta_M$ as a function of $\evh$ at fixed $V_0/V=1.8,1.7,1.6,1.4$ from the top line to the bottom line.}
    \end{center}
%    \vskip-0.5cm
\end{figure}

A microscopic theory is then developed to show that the unusually large band structure and FS renormalization, the absence of magnetism, and the emergence of the $d$-wave nematic order in FeSe can be explained by the important Fe-Fe interatomic Coulomb repulsion $V$. In addition to offering a natural description of the ARPES and quantum oscillation experiments, the theory suggests that the electronic nematicity is the driving force behind the tetragonal to orthorhombic structural transition. Interestingly, recent neutron scattering experiments in FeSe observed the spin fluctuations around $(\pi,0)$ below the structural transition and the magnetic resonance at $(\pi,0)$ below the superconducting transition \cite{boothroyd,jzhao}, which can be explained by the presence of a nematic electronic structure \cite{kreisel}. We also note that the same $V$-term has been argued recently to play an important role in stabilizing the $s$-wave pairing symmetry in Fe-based superconductors \cite{hu2015}. The overarching importance of the extended Coulomb interaction $V$ may originate from the lack of the charge reservoir layers and the shorter Fe-Fe bond in bulk FeSe when compared to Fe-pnictides \cite{eschrig}. Interestingly, electronic nematicity with similar phenomenology has been observed recently in $35$-monolayer FeSe films with a larger $\Delta_M=80$meV and higher $T_{\rm nem}=125$K \cite{dhlu,feng}, suggesting that further reduced screening of extended Coulomb interaction in films can result in a stronger $V$ and an enhanced nematic response.

An important, falsifiable prediction of the present theory is the correlation between the emergence of the degeneracy lifting nematic state and the dynamical, inter-site Coulomb interaction $V$-induced proximity of the vHS to the Fermi level. It is thus desirable to seek direct experimental evidence for extended Coulomb interaction and further experimental tests by other techniques such as scanning probe and X-ray spectroscopy for the presence of the vHS near $E_F$ above and its removal below the nematic/structural transition. This state of affairs is summarized in Fig.~6 where $\Delta_M$ is plotted versus the distance of the vHS to the Fermi level $\evh$ at different values of $V_0$. It is in principle possible to tune the vHS by doping, pressure, or chemical substitution and study the corresponding changes in the nematic response such as the transition temperature and the band degeneracy splitting energy $\Delta_M$. While more experimental tests are clearly necessary, recent studies of chemically substituted bulk FeSe by S indeed find that $\Delta_M$ increases with decreasing $\evh$ \cite{watson15}, qualitatively consistent with the prediction shown in Fig.~6(c).

More importantly, since electron doping FeSe moves the Fermi level upward and away from the vHS at $M$ point (see Fig.~6a), the present theory predicts that the nematic order will disappear while the vHS and band degeneracies survive at low temperatures when the material is subject to sufficient electron doping. Remarkably, this has been observed recently by ARPES on bulk FeSe whose surface layer is heavily electron-doped with Na \cite{kim}. The measurements show that the nematic state is absent and the vHS and the band degeneracy remain intact at $65$meV below the Fermi level \cite{kim}. Surprisingly, a pairing gap near $E_F$ develops at low temperatures that is consistent with the onset of a superconducting transition at $T_c=20$K, much higher than the $9$K transition in undoped FeSe, suggesting that the nematic state in bulk FeSe is a form of competing order of the superconducting state. Furthermore, in $50$-monolayer FeSe films, a continuous reduction of the nematic order induced degeneracy splitting $\Delta_M$ by surface electron doping with K has been observed, as well as an increase in the superconducting $T_c$ when nematicity is suppressed \cite{chwen}.

There is indeed an empirical correlation that the higher the {\em electronic} nematic transition temperature, the higher the optimal superconducting $T_c$ when electron doping removes the nematicity. Recent reports on $30$-monolayer FeSe films, which are similar to the $35$-monolayer films mentioned earlier with a nematic transition temperature around $125$K \cite{dhlu,feng}, shows a superconducting $T_c$ as high as $44$K \cite{chwen} under K surface doping. The strongest nematic phase with $T_{\rm nem}\simeq180$K is in fact an insulator observed in the nonsuperconducting N-phase of single and double-layer FeSe films grown on SrTiO$_3$ substrates \cite{zhou13,zhou15}. Thermal annealing introduces significant electron doping that removes the nematic insulating state in favor of the superconducting S-phase with the highest $T_c$ as much as $65$K \cite{xue,zhou13,zhou15}.
%It is conceivable that the intersite Coulomb $V$ discussed here plays an important role in the emergence of the insulating nematic N-phase.
Our findings on the importance of inter-site Coulomb interaction, the correlation induced proximity of the vHS near the Fermi level, and the new form of nonlocal, bond nematic orbital order with momentum space anisotropy provide considerable new microscopic insights into the intimate, competing relationship between nematicity and superconductivity, which may hold the key to understanding the pairing mechanism and to making $T_c$ even higher in these materials.

%Whether the latter belongs to the $d$-wave nematic bond ordered state and its relation to the superconducting S-phase in the single-layer FeSe are interesting questions for future study.

\section{Acknowledgement}

We thank Peng Zhang, Sen Zhou and Junfeng He for helpful discussions. This work was supported by the U.S. Department of Energy, Office of Science, Basic Energy Sciences, under Award DE-FG02-99ER45747. Z.W. thanks the Aspen Center for Physics for hospitality and the support of ACP NSF grant PHY-1066293.


\begin{thebibliography}{99}
\bibitem{footnote}
Breaking the glide symmetry would introduce the mixing of $g_3$ and $g_{5+}^{\prime\prime}$ type of interactions that generates the dimerization between Fe$_A$ and Fe$_B$ as observed recently by electron diffraction \cite{mcqueen}.
\end{thebibliography}
\end{document}